\documentclass[twocolumn,pra,aps]{revtex4-1}

\usepackage[dvipdfmx]{graphicx}
\usepackage{color, fancybox} 
\usepackage{amsfonts}
\usepackage{amssymb,amsmath,amsthm} 
\usepackage{braket}

\begin{document}
\newcommand{\mbold}[1]{\mbox{\boldmath $#1$}}
\newcommand{\sbold}[1]{\mbox{\boldmath ${\scriptstyle #1}$}}
\newcommand{\tr}{\,{\rm tr}\,}
\title{Minimum-uncertainty states and completeness of non-negative quasi probability of finite-dimensional quantum systems}
\author{T.~Hashimoto, A.~Hayashi, and M.~Horibe}
\affiliation{Department of Applied Physics, 
          University of Fukui, Fukui 910-8507, Japan}
\begin{abstract}
We construct minimum-uncertainty states and a non-negative quasi probability distribution for quantum systems on a finite-dimensional space. 
We reexamine the theorem of Massar and Spindel for the uncertainty relation
of the two unitary operators related by the discrete Fourier transformation. 
It is shown that some assumptions in their proof can be justified by the 
use of the Perron-Frobenius theorem.  
The minimum-uncertainty states are the ones that saturate this uncertainty inequality. The continuum limit is closely analyzed by introducing a scale 
factor in the limiting scheme. 
Using the minimum-uncertainty states, we construct a non-negative quasi probability distribution. Its marginal distributions are smeared out. 
However, we show that this quasi probability is optimal in the sense that 
there does not exist a non-negative quasi probability distribution with 
sharper marginal properties if the translational covariance in the phase 
space is assumed. 
Generally, it is desirable that the quasi probability is complete, i.e., 
it contains full information of the state.  
We show that the obtained quasi probability is indeed complete if the dimension of the state space is odd, whereas it is not if the dimension is even.
\end{abstract}

\pacs{PACS:03.67.Hk}
\maketitle
\section{Introduction}
The uncertainty principle \cite{Heisenberg1927} is arguably one of the 
most fundamental features that differentiate quantum mechanics  from 
classical mechanics.  
It states that the product of uncertainties in complementary physical observables 
(e.g. position and momentum) has an inherent finite lower bound, and it 
has a profound influence on our view of the physical world. 
Because of the uncertainty principle, the dynamics of a quantum system is qualitatively 
different from a classical one; for example, an atom would collapse 
without this principle.  
Furthermore, recent studies show that the uncertainty principle 
also plays an important role in a variety types of  quantum information processings 
\cite{Nielsen_text_book}. 
For example, quantum cryptography \cite{BB84}, one of the most remarkable applications of quantum information, exploits the uncertainty principle together with 
the no-cloning theorem \cite{Wootters82} to ensure its provable security.

The uncertainty relation of the position and momentum in the continuous 
quantum mechanics is expressed by an inequality involving  
the standard deviations of their distributions \cite{Kennard1927}; 
that is, $\Delta x \Delta p \ge 1/2$. The states that attain the minimum  are called minimum-uncertainty states, and they are given by the 
coherent states. 
The coherent states, the eigenstates of the annihilation operator, have interesting 
properties and useful applications in various fields of physics 
(see, e.g., Ref.~\cite{Klauder1985}).  Using the coherent states, one 
can define a quasi probability distribution for the position and momentum variables, 
which is called the Husimi function (Q-distribution) \cite{Husimi1940}. 
The Husimi function is always non-negative in contrast to the Wigner 
function \cite{Wigner1932}, which is another quasi distribution function and may 
take negative values except for the case of Gaussian wave functions 
\cite{Hudson1974}. 

In this paper, we study analogous minimum-uncertainty states and 
a non-negative quasi probability distribution for finite-dimensional 
quantum systems (qudits). To define the position and momentum coordinates, 
we take two bases related by the discrete Fourier transformation.  
The modulus of the expectation value of the position (momentum)  translation operator is suitable for quantifying the uncertainty of 
the position (momentum) distribution \cite{Opatrny95, Opatrny96,Massar08}. 
For other approaches using the Jacobi theta function to 
construct analogous minimum-uncertainty states for a qudit, 
see, e.g., \cite{Klimov09, Cotfas12, Marchiolli12}.

Massar and Spindel derived an inequality for 
the expectation values of the above two translation operators 
(Theorem 2 in \cite{Massar08}). They also
discussed the minimum-uncertainty states saturating their inequality 
(Theorem 3 in \cite{Massar08}), which involves two 
assumptions for the greatest eigenvalue and the associated eigenvector 
of the Harper operator. 
We will show that these two assumptions can be justified using the 
Perron-Frobenius theorem (see, e.g.,\cite{Horn1985}), and we provide a 
detailed proof of a theorem combining those of Massar and Spindel 
(Sec.~\ref{sec_Minimum}).

We call the states saturating this inequality minimum-uncertainty states, 
which comprise an overcomplete set in the state space. 
In Sec.~\ref{sec_continuum}, we will give a close analysis to show
that these minimum-uncertainty states approach the coherent states as the 
dimension of the state space goes to infinity. 

In the same way as in continuous quantum mechanics, we define a quasi 
probability distribution of a qudit using the minimum-uncertainty states 
(Sec.~\ref{sec_non-negative}). 
This is a finite-dimensional version of the Husimi function, and 
non-negative at the cost of the smeared out marginal distributions.    
We show that the obtained quasi probability distribution is optimal in the sense 
that there exists no non-negative quasi probability distribution 
with sharper marginal properties if the translational covariance is assumed. 

In continuous quantum mechanics, the Husimi function is complete, i.e.,  
it contains full information of the state. This is one of the desirable 
properties of quasi probabilities of quantum systems.
For finite-dimensional quantum systems, however, we find that the obtained 
quasi probability is indeed complete if the dimension of the state space is 
odd, whereas it is not if the dimension is even (Sec.~\ref{sec_completeness}).

 
\section{Minimum-uncertainty states of a finite-dimensional quantum system} 
\subsection{Position and momentum uncertainty of a qudit}
\label{sec_uncertainty}
We consider a qudit, a quantum system described  by a $d$-dimensional 
complex linear space $\mathbb{C}^d$. 
An orthonormal basis $\{\ket{a}\}_{a=0}^{d-1}$ 
is fixed to define  the ``position'' coordinate $a$. We introduce another  orthonormal 
basis, which is the discrete Fourier transform defined by 
\begin{align}
 \ket{\tilde b} = \frac{1}{\sqrt{d}}\sum_{a=0}^{d-1} \omega^{ba} \ket{a}
,\ b=0,1,\ldots,\ldots,d-1,
\end{align}
where $\omega=e^{2\pi i/d}$ is a primitive $d$th root of unity. 
The index $b$ is interpreted as the ``momentum'' coordinate. 
These two bases are unbiased in the sense that 
$|\braket{a|\tilde b}|=1/\sqrt{d}$ for all $a$ and $b$, and they are 
expected to approach the continuous position and momentum bases as 
the dimension $d$ goes to infinity. 
As a feature of the discrete Fourier transform, the position and momentum 
coordinates, $a$ and $b$, can not simultaneously have sharp values.  
 
To quantify the uncertainty with respect to these two bases, we 
employ  two unitary operators $Q$ and $P$. The operator $Q$ 
is given by
\begin{align}
 Q = \sum_{a=0}^{d-1} \omega^a \ket{a}\bra{a}, 
\end{align}
which is diagonal in the position basis $\{\ket{a}\}$. 
In the momentum basis $\{\ket{\tilde b}\}$, the operator $Q$ translationally 
shifts the momentum coordinate as $Q\ket{\tilde b} = \ket{\widetilde{ b+1}}$.
Here, it is assumed that  if $b+1=d$ then $\ket{\widetilde{ b+1}}$ is equal 
to $\ket{\tilde 0}$. 
Throughout this paper we employ this periodic convention for the position and 
momentum coordinates; namely, we assume that 
\begin{align}
  \ket{a+d} = \ket{a},\ \ket{\widetilde{b+d}} = \ket{\tilde b}, 
\end{align}
for any integers $a$ and $b$. Another operator $P$ is defined by 
\begin{align}
 P  = \sum_{b=0}^{d-1} \omega^{-b} \ket{\tilde b}\bra{\tilde b}, 
\end{align}
which is diagonal in the momentum basis, and in the position basis it acts as 
the translational operator; $ P\ket{a} = \ket{a+1}$. 
It is readily shown that $P$ and $Q$ satisfy the following relations:
\begin{align}
  Q^d  = P^d = \mbold{1},\ QP = \omega PQ.
\end{align}
The relation $QP = \omega PQ$ can be regarded as  the counterpart of 
the canonical commutation relation of  the continuous position and momentum operators. 

For a general state $\ket{\phi}$, we write 
\begin{align}
  \ket{\phi} = \sum_{a=0}^{d-1} c_a \ket{a} 
                    = \sum_{b=0}^{d-1}  \tilde c_b \ket{\tilde b}, 
\end{align}
where $c_a$ and $\tilde c_b$ are expansion coefficients in the position 
and momentum basis, respectively.  
Then the expectation values of $Q$ and $P$ for the state $\ket{\phi}$ 
take the following form:
\begin{align}
  &\braket{\phi|Q|\phi} = \sum_{a=0}^{d-1} |c_a|^2 \omega^a
                                        = \sum_{b=0}^{d-1} \tilde c_{b+1}^* \tilde c_b, 
                                                        \label{eq_Qexp} \\
  &\braket{\phi|P|\phi} = \sum_{a=0}^{d-1} c_{a+1}^*c_a
                                      = \sum_{b=0}^{d-1} |\tilde c_b|^2\omega^{-b}.
                                                        \label{eq_Pexp}
\end{align}

\begin{figure}
\includegraphics[width=7cm]{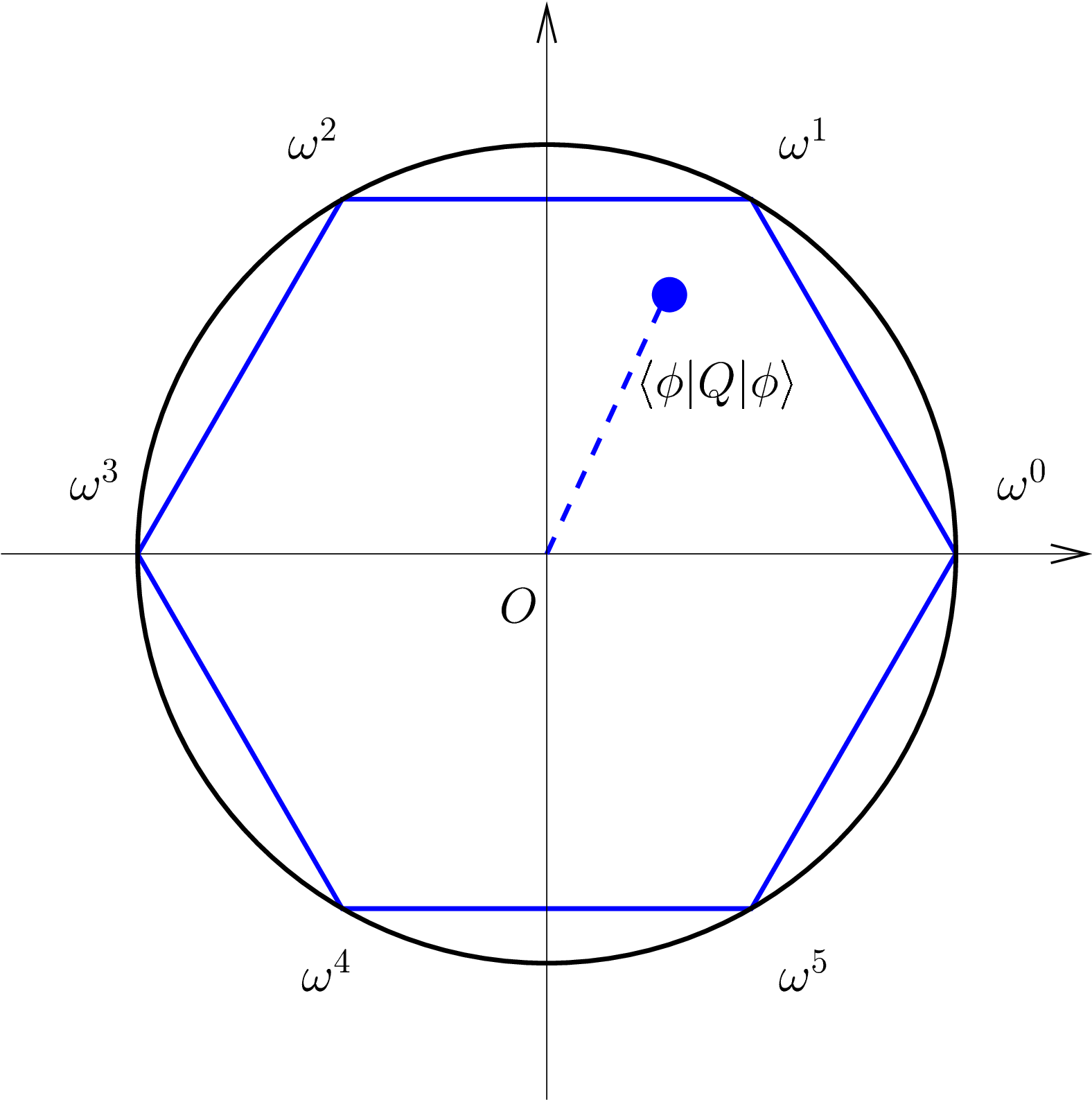}
\caption{(Color online) The $d$th roots of unity in the complex plane and the expectation 
value $\braket{\phi|Q|\phi}$ represented by a point in the regular $d$-sided 
polygon formed by these roots. This figure displays the case of $d=6$.}
\label{fig:hull}
\end{figure}

Now let us examine the expectation value $\braket{\phi|Q|\phi}$ expressed in terms 
of $c_a$.  This is an average of roots of unity $\omega^a$ with weights given by $|c_a|^2$. 
In the complex plane, the points $\{\omega^a\}_{a=0}^{d-1}$ are 
at the vertices of a regular $d$-sided polygon inscribed in the unit circle, and 
the expectation value $\braket{\phi|Q|\phi}$ is somewhere in this 
polygon (see Fig.~\ref{fig:hull}). If the position coordinate has a sharp 
value, say $a_0$, $\braket{\phi|Q|\phi}$ is at the vertex $\omega^{a_0}$. 
In this case, and only in this case, $|\braket{\phi|Q|\phi}|$ is equal to 1, 
otherwise we generally have $|\braket{\phi|Q|\phi}| < 1$. 
In contrast, if the weight is equally distributed as $|c_a|^2=1/d$, 
$\braket{\phi|Q|\phi}$ is at the origin; that is, $|\braket{\phi|Q|\phi}|=0$. 
Thus the quantity $|\braket{\phi|Q|\phi}|$ is a measure of quantifying how 
sharply the position coordinate is distributed. In the same way the quantity 
 $|\braket{\phi|P|\phi}|$ measures the sharpness of the distribution of 
momentum coordinate. However, the quantities $|\braket{\phi|Q|\phi}|$ 
and $|\braket{\phi|P|\phi}|$ cannot simultaneously have their maximum value 1. 
For example, take the case of $|\braket{\phi|Q|\phi}|=1$ which occurs only when 
$|c_a|$ is nonzero for a certain single value of $a$. In this case, however, 
 $|\braket{\phi|P|\phi}|$ must be 0, as its expression in terms of $c_a$ 
clearly shows.

Motivated by these considerations, we define the certainty $C$ of a state 
$\ket{\phi}$ to be 
\begin{align}
   C(\ket{\phi}) = |\braket{\phi|Q|\phi}\braket{\phi|P|\phi}|, 
                 \label{eq_certainty}
\end{align}
to quantify the mutual uncertainty with respect to the position and momentum 
coordinates. Note that a larger $C$ means less uncertainty as the name ``certainty'' 
indicates. 

\subsection{Minimum-uncertainty states}  \label{sec_Minimum}
In the preceding section, we have seen that the certainty $C(\ket{\phi})$
in Eq.~(\ref{eq_certainty}) 
serves as a measure of certainty of position and momentum for a qudit state $\ket{\phi}$.
In this section we study the maximum value of the certainty and the states 
attaining the maximum certainty: the states with the minimum uncertainty. 

Let us first examine the case of a qubit, $d=2$. 
In the 2-dimensional case,  the operators 
$Q$ and $P$ are given by the Pauli matrices,  
\begin{align}
  Q &= \ket{0}\bra{0}-\ket{1}\bra{1} = \sigma_z, 
                        \nonumber \\
  P &=  \ket{0}\bra{1}+\ket{1}\bra{0} = \sigma_x.
                       \nonumber
\end{align}
The states are conveniently expressed by the Bloch vector representation, 
\begin{align}
   \ket{\mbold{n}} = \cos \frac{\theta}{2} \ket{0} 
                                  + e^{i\varphi}\sin \frac{\theta}{2} \ket{1}, 
\end{align}
where $\mbold{n} = (\sin\theta\cos\varphi, \sin\theta\sin\varphi, \cos\theta)$ is 
the Bloch vector. For the certainty $C$ of the state $\ket{\mbold{n}}$, we obtain 
\begin{align}
   C(\mbold{n}) = |\braket{\mbold{n}|\sigma_z|\mbold{n}}
                                \braket{\mbold{n}|\sigma_x|\mbold{n}}|
                           = |n_z n_x|.
\end{align}
The upper bound of $C(\mbold{n})$ is readily determined by using the 
following inequalities:
\begin{align}
   |n_z n_x| \le \frac{n^2_x+n^2_z}{2} \le \frac{1}{2}.
\end{align}
Thus the maximum value of the certainty $C$ is $1/2$, and the maximum is attained 
by the following four Bloch vectors:
\begin{align}
   \mbold{n}^{(\alpha,\beta)} = 
               \frac{1}{\sqrt{2}} 
      \left( (-1)^\alpha, 0, (-1)^\beta \right),\  (\alpha,\beta=0,1).
                    \label{eq_Bloch}
\end{align}  
The state with $ \mbold{n}^{(0,0)}$ is denoted by 
$\ket{\Gamma}$, and it takes the following explicit form: 
\begin{align}
  \ket{\Gamma} &\equiv \cos\frac{\pi}{8} \ket{0} + \sin\frac{\pi}{8} \ket{1}
                                    \nonumber \\
                           &= \frac{\sqrt{2+\sqrt{2}}}{2}\ket{0}
                               +\frac{\sqrt{2-\sqrt{2}}}{2}\ket{1}.
\end{align}
It should be noticed that the four states attaining the maximum $C$ can be 
expressed as
\begin{align}
   \ket{\alpha,\beta} \equiv 
         \sigma_x^\alpha \sigma_z^\beta \ket{\Gamma},\  (\alpha,\beta=0,1).
\end{align} 

Now we generalize these results to arbitrary-dimensional cases, and 
establish the following theorem:

\smallskip
{\bf Theorem}: For any normalized state $\ket{\phi}$,
\begin{itemize} 
\item[(i)] The certainty $C$ is bounded by the inequality,
\begin{align}
    C(\ket{\phi}) 
    \equiv |\braket{\phi|Q|\phi}\braket{\phi|P|\phi}| \le h^2,
            \label{eq_main_inequality}
\end{align}
where $h$ is the greatest eigenvalue of Harper operator $H$ given by 
\begin{align}
   H \equiv (P + P^{\dagger} + Q + Q^{\dagger})/4.
           \label{eq_H}
\end{align}
\item[(ii)] Equality in (\ref{eq_main_inequality}) holds if and only if 
\begin{align}
   \ket{\phi} = P^{\alpha}Q^\beta \ket{\Gamma}\ (\text{up to a global phase}),
\end{align}
where $\ket{\Gamma}$ is the nondegenerate eigenstate of $H$ with the greatest eigenvalue
 $h$, and $\alpha$ and $\beta$ are integers ($\alpha,\beta=0,1,\ldots,d-1$). 
The states $\ket{\alpha,\beta} \equiv P^\alpha Q^\beta \ket{\Gamma}$ are called 
the minimum-uncertainty states.
\end{itemize}
Statement (i) is essentially a special case ($\theta = \pi/4$) of theorem 2  
shown by Massar and Spindel in \cite{Massar08}. 
For later convenience, we give its proof below. 
Statement (ii) corresponds to theorem 3 in \cite{Massar08}, which was proved  
by assuming that the greatest eigenvalue $h$ of $H$ is 
nondegenerate and the associated eigenstate satisfies $\braket{a|\Gamma} \ne 0$.
We will show that these two assumptions can be justified using the 
Perron-Frobenius theorem, which will also be powerful when we later discuss 
the completeness of the quasi probability.  
For an analysis of the eigenstructure of the Harper operator in terms 
of the crossing number, see \cite{Barker00}.  

\subsubsection{Proof of statement (i) in Theorem}
In order to prove statement (i) in Theorem, we start with an inequality,  
\begin{align}
  \sqrt{|\braket{\phi|Q|\phi}\braket{\phi|P|\phi}|} \le 
    \frac{1}{2} \left( |\braket{\phi|Q|\phi}|+|\braket{\phi|P|\phi}| \right), 
              \label{eq_arithmetic}
\end{align}
where equality holds if and only if $|\braket{\phi|Q|\phi}|=|\braket{\phi|P|\phi}|$. 

We write a given state $\ket{\phi}$ in the basis $\{\ket{a}\}$ as  
\begin{align}
   \ket{\phi}= \sum_{a=0}^{d-1}c_a \ket{a}. 
\end{align}
Replacing  expansion coefficients $c_a$ by their moduli $|c_a|$, we introduce 
a new state $\ket{\phi'}$  as
\begin{align}
  \ket{\phi'}= \sum_{a=0}^{d-1}|c_a| \ket{a}
                    = \sum_{b=0}^{d-1} \tilde c'_b \ket{\tilde b},
\end{align}
where expansion coefficients of $\ket{\phi'}$ in the basis $\{\ket{\tilde b}\}$ are 
denoted by $\tilde c'_b$. We further define another state $\ket{\phi''}$ by replacing 
$\tilde c'_b$ by  $|\tilde c'_b|$, that is,  
\begin{align}
  \ket{\phi''}= \sum_{b=0}^{d-1} |\tilde c'_b| \ket{\tilde b}.
\end{align}
Using Eqs.~(\ref{eq_Qexp}) and (\ref{eq_Pexp}), we can readily show that the following 
relations hold:
\begin{subequations}
\begin{align}
               &\braket{\phi'|P|\phi'} \ge  |\braket{\phi|P|\phi}|, 
                         \label{eq_PQphi1a} \\
               &\braket{\phi'|Q|\phi'} = \braket{\phi|Q|\phi},         
                         \label{eq_PQphi1b}
\end{align}
\end{subequations}
and
\begin{subequations}
\begin{align}
               &\braket{\phi''|P|\phi''} =  \braket{\phi'|P|\phi'}, 
                                \label{eq_PQphi2a}\\
               &\braket{\phi''|Q|\phi''} \ge |\braket{\phi'|Q|\phi'}|.   
                               \label{eq_PQphi2b}
\end{align}
\end{subequations}
Note that $\braket{\phi''|P|\phi''}$ and $\braket{\phi''|Q|\phi''}$ are real, 
and therefore, $\braket{\phi''|P|\phi''}=\braket{\phi''|P^\dagger|\phi''}$ 
and $\braket{\phi''|Q|\phi''}=\braket{\phi''|Q^\dagger|\phi''}$. 
Thus we have 
\begin{align}
  \frac{|\braket{\phi|Q|\phi}|+|\braket{\phi|P|\phi}|}{2}
 \le \bra{\phi''}\frac{P+P^\dagger+Q+Q^\dagger}{4}\ket{\phi''}.
\end{align}
The right-hand side is clearly less than or equal to $h$, the greatest eigenvalue of $H$, 
\begin{align}
   \bra{\phi''}\frac{P+P^\dagger+Q+Q^\dagger}{4}\ket{\phi''} \le h.
             \label{eq_lastinequality}
\end{align}
Combining this result and inequality (\ref{eq_arithmetic}), we obtain inequality (\ref{eq_main_inequality}). 

\subsubsection{Eigenstate of $H$ with the greatest eigenvalue} \label{subsec_H}
Before proving statement (ii) of Theorem, we study the properties of the eigenstate of $H$ 
with the greatest eigenvalue $h$. Some of them will be needed in the proof of statement (ii). 
We will show the following:
\begin{itemize}
\item[(a)] The greatest eigenvalue $h$ is positive and not degenerate. The phase of  corresponding 
eigenstate $\ket{\Gamma}$ can be chosen such that $\braket{a|\Gamma}$ is 
real and strictly 
positive for all $a$. 
\item[(b)] The eigenstate $\ket{\Gamma}$ is invariant under the Fourier transformation; 
$F\ket{\Gamma}=\ket{\Gamma}$, where $F$ is the Fourier transform 
operator defined by
\begin{align}
    F = \sum_{a=0}^{d-1}\ket{\tilde a}\bra{a}, 
\end{align}
and, hence $\braket{a|\Gamma}=\braket{\tilde a|\Gamma}=\braket{-a|\Gamma}$.
\item[(c)] The following relations hold:
\begin{align} 
   h&=\braket{\Gamma|Q|\Gamma} = \braket{\Gamma|Q^{\dagger}|\Gamma}
                       \nonumber \\
     &=\braket{\Gamma|P|\Gamma} = \braket{\Gamma|P^{\dagger}|\Gamma}.      
                      \label{eq_PQH}  
\end{align}
\end{itemize}

To show that the above statement (a) holds, some known properties of elementwise 
positive matrices will be employed. 
Here, we treat operators in the matrix representation based on the basis 
$\{\ket{a}\}_{a=0}^{d-1}$.
We introduce a real symmetric matrix $H_\kappa \equiv H+\kappa \mbold{1}$ 
with $\kappa$ a real number. The off-diagonal part of $H_\kappa$ is 
given by $(P+P^\dagger)/4$, all of whose elements are non-negative. 
The diagonal part, $(Q+Q^\dagger)/4+\kappa \mbold{1}$, is denoted $D$, and 
all of its diagonal elements are strictly positive for a sufficiently 
large $\kappa$.  
Now consider $H_\kappa^{d-1}$ and expand it in terms of $P$, $P^\dagger$, 
and $D$.
For any $i \le j$, there is a term of the form 
$(P^\dagger/4)^{j-i}D^{d-1-j+i}$ that has a strictly positive $(i,j)$ entry while 
other terms are elementwise non-negative. 
For the $(j,i)$ entry, a similar argument can be applied. 
Thus the matrix $H_\kappa^{d-1}$ is elementwise strictly positive. 

Now recall that, according to the Perron-Frobenius theorem, the eigenvalue of the 
largest modulus of an elementwise strictly positive matrix is real and 
nondegenerate, and  the associated eigenvector can be chosen to have 
strictly positive components (see, e.g., \cite{Horn1985}) . 
The eigenvalues of $H_\kappa^{d-1}$ are clearly given by 
$(\kappa+\lambda_i)^{d-1}$, with $\lambda_i$ being real eigenvalues of $H$. 
Thus we conclude that the greatest eigenvalue of $H$  is not degenerate 
and the associated eigenstate $\ket{\Gamma}$ can be chosen so that 
$\braket{a|\Gamma}>0$ for all $a$. 

To show that $h>0$, note that  the trace of $H$ is 0. 
In the case of $d>1$, this is possible only when $h>0$ since $h$ is the 
unique greatest eigenvalue. When $d=1$, it is evident that $h=1$. 

Now we show that $F\ket{\Gamma}=\ket{\Gamma}$. 
It is easy to show that $FQF^\dagger = P^\dagger$ and $FPF^\dagger = Q$, 
and hence $H$ commutes with $F$.  
This implies that $\ket{\Gamma}$ is an eigenstate of $F$ since 
the greatest eigenvalue $h$ is not degenerate.  The possible eigenvalues of $F$ are 
1, $-1$, $i$, and $-i$. This is because $F^2 = T$, where $T$ is the reflection operator given by
\begin{align}
    T = \sum_{a=0}^{d-1} \ket{-a}\bra{a},
\end{align}
and $T$ satisfies  $T^2= \mbold{1}$.
Assume that $F\ket{\Gamma}=f\ket{\Gamma}$ with $f$ being 1, -1, $i$, or $-i$. 
This is explicitly written as 
\begin{align}
   \sum_{a'=0}^{d-1} \braket{a|F|a'} \braket{a'|\Gamma} = f\braket{a|\Gamma},
\end{align}
where $ \braket{a|F|a'} =\omega^{aa'}/\sqrt{d}$. Setting $a=0$, we observe  
\begin{align}
    \frac{1}{\sqrt{d}}\sum_{a'=0}^{d-1} \braket{a'|\Gamma} = f \braket{0|\Gamma}.
\end{align}
This requires that $f=1$ since $\braket{a|\Gamma} > 0$ for all $a$.  
From $F\ket{\Gamma}=\ket{\Gamma}$, it immediately follows that  
$\braket{a|\Gamma}=\braket{\tilde a|\Gamma}=\braket{-a|\Gamma}$.

Further, the invariance $F\ket{\Gamma}=\ket{\Gamma}$ implies 
\begin{align}
   \braket{\Gamma|Q|\Gamma} = \braket{\Gamma|F^\dagger QF|\Gamma} = \braket{\Gamma|P|\Gamma}.
\end{align}

Since $\braket{a|\Gamma}$ and $\braket{\tilde b|\Gamma}$ are real,  
 $\braket{\Gamma|P|\Gamma}$ and $\braket{\Gamma|Q|\Gamma}$ are 
also real. We therefore find  
\begin{align} 
  \braket{\Gamma|Q|\Gamma} = \braket{\Gamma|Q^{\dagger}|\Gamma}
 =\braket{\Gamma|P|\Gamma} = \braket{\Gamma|P^{\dagger}|\Gamma},      
\end{align}
which shows that each one is equal to $h$.
Thus we obtain Eq.~(\ref{eq_PQH}).  

Explicit analytical solutions of $h$ and $\ket{\Gamma}$ in general dimensions 
have not been obtained, but some of the results in the low-dimensional cases 
are collected in \cite{Marchiolli13}. 

\subsubsection{Proof of statement (ii) in Theorem}
``If part''  is evident. When $\ket{\phi}=P^{\alpha}Q^\beta\ket{\Gamma}$, 
we find that 
\begin{align}
  &|\braket{\phi|P|\phi}| = |\braket{\Gamma|P|\Gamma}|=h,\\ 
  &|\braket{\phi|Q|\phi}| = |\braket{\Gamma|Q|\Gamma}|=h,
\end{align}
which shows that  $\ket{\phi}$ satisfies the equality in (\ref{eq_main_inequality}).

 Proving  ``only if part'' is rather involved. 
Suppose that $\ket{\phi}$ satisfies the equality in (\ref{eq_main_inequality}).  
In the same way as in the proof of statement (i), we define 
$\ket{\phi'}$ and $\ket{\phi''}$ as follows:
\begin{align}
  \ket{\phi} &= \sum_{a=0}^{d-1}c_a \ket{a}, \\ 
  \ket{\phi'} &= \sum_{a=0}^{d-1}|c_a| \ket{a}
                    = \sum_{b=0}^{d-1} \tilde c'_b \ket{\tilde b}, \\
  \ket{\phi''}&= \sum_{b=0}^{d-1} |\tilde c'_b| \ket{\tilde b}.
\end{align}
This time the equality should hold in all inequalities in the proof of statement (i). 

First we note that 
the equality in (\ref{eq_lastinequality}) is satisfied only if $\ket{\phi''}=\ket{\Gamma}$  up to 
a global phase since the greatest eigenvalue $h$ is not degenerate.  

Second we examine the equality in  (\ref{eq_PQphi2b}), 
$\braket{\phi''|Q|\phi''}= |\braket{\phi'|Q|\phi'} |$. This  is explicitly written as 
\begin{align}
   \sum_{b=0}^{d-1} |\tilde c'_{b+1}||\tilde c'_b| = 
                \left|\sum_{b=0}^{d-1} \tilde c_{b+1}^{'*}\tilde c'_b \right|, 
\end{align}
which implies that all terms on the right-hand side must have the same phase factor, that is, 
$\tilde c_{b+1}^{'*}\tilde c'_b= |\tilde c_{b+1}^{'}\tilde c'_b|u$ 
, with $u$ being a complex number of unit modulus.
This relation can be rewritten as
\begin{align}
    \frac{\tilde c'_b}{|\tilde c'_b|}=u \frac{\tilde c'_{b+1}}{|\tilde c'_{b+1}|}, 
    \ (b=0,1,\ldots,d-1),
\end{align}
Note that  $|\tilde c'_b| > 0$ for all $b$ since $\ket{\phi''}=\ket{\Gamma}$, 
and the above relation is well defined.
Using this relation successively we obtain
\begin{align}
     \frac{\tilde c'_b}{|\tilde c'_b|} = u^{-b}\frac{\tilde c'_0}{|\tilde c'_0|}.
\end{align}
Setting $b=d$ and remembering $\tilde c'_d=\tilde c'_0$ by our convention,  
we find that the phase factor $u$ must be a $d$th root of unity, $\omega^\alpha$ 
with some integer $\alpha$. Thus the $b$ dependence of the phase of $\tilde c'_b$ 
is given by $\omega^{-\alpha b}$, from which we conclude that  
$\ket{\phi'}=P^{\alpha}\ket{\phi''}=P^{\alpha}\ket{\Gamma}$ up to a global phase. 

Let us now turn to the equality in (\ref{eq_PQphi1a}), 
$\braket{\phi'|P|\phi'} =  |\braket{\phi|P|\phi}|$, which is explicitly written as
 \begin{align}
   \sum_{a=0}^{d-1} |c_{a+1}||c_a| = 
                \left|\sum_{a=0}^{d-1} c_{a+1}^{*}c_a \right|, 
\end{align}
Since $\ket{\phi'}=P^{\alpha}\ket{\Gamma}$, we have $|c_a|>0$ for all $a$.
We can repeat a similar argument to the preceding one, and we find that $\ket{\phi}$ 
is given by $Q^\beta\ket{\phi'}$ with some integer $\beta$. Combining this and the previous result, 
$\ket{\phi'}=P^{\alpha}\ket{\Gamma}$, we finally conclude that 
$\ket{\phi} = P^{\alpha}Q^\beta\ket{\Gamma}$  up to a global phase. 

It should be noted that we used the fact that 
$\braket{a|\Gamma} \ne 0$ and $\braket{\tilde b|\Gamma} \ne 0$ 
for all $a$ and $b$ in the above argumentation.

\subsubsection{Parameter $\theta$ in the theorem of Massar and Spindel}
Theorems 2 and 3 in \cite{Massar08} involve a parameter 
$\theta \in [0,\pi/2]$. They state that for any $\ket{\phi}$
the following inequality holds: 
\begin{align}
  \cos \theta |\braket{\phi|Q|\phi}|
 +\sin \theta |\braket{\phi|P|\phi}| \le h_\theta, 
                 \label{eq_MS_theorem1}
\end{align}
where $h_\theta$ is the greatest eigenvalue of the Hermitian operator 
\begin{align}
  H_{\theta} = \cos \theta \frac{P+P^\dagger}{2} 
              +\sin \theta \frac{Q+Q^\dagger}{2},          
\end{align}
and the equality in Eq.~(\ref{eq_MS_theorem1}) holds if and only if 
\begin{align}
   \ket{\phi} = P^{\alpha}Q^\beta \ket{\Gamma_\theta}
     \ (\text{up to a global phase}),
\end{align}
where $\ket{\Gamma_\theta}$ is the nondegenerate eigenstate of 
$H_\theta$ with the eigenvalue $h_\theta$. 
In this paper, we have concentrated on the case $\theta=\pi/4$.   
It is, however, clear that the two assumptions in the proof of 
``only if part'' can be justified as in the case $\theta=\pi/4$. 
This is because, except for the trivial cases  
$\theta = 0$ or $\theta = \pi/2$,  
the matrix $(H_\theta + \kappa \mbold{1})^{d-1}$ can be shown elementwise 
to be strictly positive, and therefore $h_\theta$ is nondegenerate, 
$\braket{a|\Gamma_\theta} \ne 0$, and 
$\braket{\tilde b|\Gamma_\theta} \ne 0$.

\section{Continuum limit} \label{sec_continuum}
In the continuous quantum mechanics, the minimum-uncertainty states are given 
by coherent states, which are eigenstates of the annihilation 
operator, and by translationally shifting the ground state of a 
harmonic oscillator in the phase space. 
The minimum-uncertainty state $\ket{\Gamma}$ is expected to approach 
a coherent state as the dimension $d$ goes to infinity   
(see, e.g., \cite{Massar08, Barker00}). 
The coherent states, however, may have any width, and they are all 
minimum-uncertainty states in continuous quantum mechanics.  
In this section, introducing a scale factor in the limiting scheme, 
we show how the single state $\ket{\Gamma}$ approaches coherent 
states with different widths.  

We start by writing the eigen equation $H \ket{\phi} = \lambda \ket{\phi}$ in 
the position basis $\{ \ket{a} \}$. 
\begin{align}
  \frac{1}{4} \left( 
     c_{a+1} + c_{a-1} + 2\cos\left(\frac{2\pi}{d}a\right) c_a 
                       \right) 
  = \lambda c_a,
        \label{eq_Mathieu1}
\end{align}
where $c_a= \braket{a|\phi}$. Dickinson and Steiglitz 
\cite{Dickinson1982} realized that this 
equation (\ref{eq_Mathieu1})  is a discrete version of the Mathieu equation by 
identifying $c_{a+1}-2c_a+c_{a-1}$ with the central 
second difference.  To extend this idea further, we consider the following limit: 
By introducing the lattice constant $\epsilon$, we define the system size 
$L=\epsilon d$.  The system size $L$ and the dimension $d$ go to infinity, 
and the lattice constant $\epsilon$ goes to zero, while 
$\sigma \equiv \sqrt{\epsilon L/(2\pi)}$ is fixed. 
It is this $\sigma$ that determines the scale of length. 
The factor $2\pi$ in the definition of $\sigma$ is just for later convenience. 
The position variable $x$ is defined by $x=a\epsilon$. 
Here the range of the discrete position index $a$ is taken to be 
$ \lfloor -(d-1)/2 \rfloor \le a \le \lfloor (d-1)/2 \rfloor$  
where the symbol $\lfloor \cdot \rfloor$ means the floor function. 
This ensures that, in the large $d$ limit, $x$ becomes a continuous variable  
ranging from $-\infty$ to $+\infty$.  
Note that in this scheme we have   
\begin{align}
   O(\epsilon^2) = O(1/L^2)=O(1/d). 
\end{align} 

Now we rewrite Eq.~(\ref{eq_Mathieu1}) as 
\begin{align}
  -\frac{1}{2}\frac{\delta^2 c_a}{\epsilon^2}
   +\frac{2}{\epsilon^2}\sin^2\left(\frac{\pi}{d}a\right) c_a 
     = \frac{2}{\epsilon^2}(1-\lambda)c_a ,
                  \label{eq_Mathieu2}
\end{align}
where $\delta^2$ is the central second difference given by
\begin{align}
   \delta^2 c_a = c_{a+1}-2c_a+c_{a-1}.
\end{align}
By introducing the wave function $\phi(x) = c_a\sqrt{\epsilon}$, we 
observe 
\begin{align}
    \frac{\delta^2 c_a}{\epsilon^2} \sqrt{\epsilon}
                              = \phi''(x) + O\left(\frac{1}{d}\right),
\end{align}
and
\begin{align}
    \frac{2}{\epsilon^2}\sin^2\left(\frac{\pi}{d}a\right)
              = \frac{x^2}{2\sigma^4} + O\left(\frac{1}{d}\right).
\end{align} 
Thus, in the leading order, Eq.~(\ref{eq_Mathieu2}) takes the form 
\begin{align}
  -\frac{1}{2}\phi''(x) + \frac{x^2}{2\sigma^4}  \phi(x) = 
         \frac{2}{\epsilon^2}(1-\lambda)\phi(x),
\end{align}
which is the Schroedinger equation of the harmonic oscillator with the 
angular frequency given by $1/\sigma^2$. 
The eigen energy of this harmonic oscillator is given by 
$(n+1/2)/\sigma^2$ where $n=0,1,\ldots$.
We thus find  
\begin{align}
   \lambda = 1-\left(n+\frac{1}{2}\right)\frac{\pi}{d},
\end{align}
from which we obtain the asymptotic expression of the greatest 
eigenvalue $h$ to be  
\begin{align}
   h = 1-\frac{\pi}{2d},\ \ (\text{as } d \rightarrow \infty).
                           \label{eq_h_asym}
\end{align}
The corresponding ground state wave function is given by a Gaussian 
function 
\begin{align}
    \phi(x) \propto \exp\left(-\frac{x^2}{2\sigma^2}\right)
                           =\exp\left(-\frac{\pi}{d}a^2 \right).
\end{align}
Thus the asymptotic form of the minimum-uncertainty state $\ket{\Gamma}$ 
is given by
\begin{align}
   \braket{a|\Gamma} = {\cal N} \exp\left(-\frac{\pi}{d}a^2\right),\ 
        (\text{as } d \rightarrow \infty),
                \label{eq_alpha_asym} 
\end{align}        
where  $ [-(d-1)/2] \le a \le [(d-1)/2]$, and  $\cal N$ is a normalization constant.

\begin{figure}
\includegraphics[width=\linewidth]{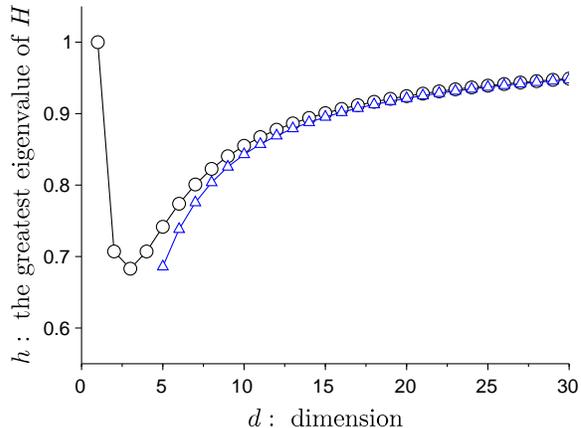}
\caption{(Color online) The greatest eigenvalue $h$ of the operator $H$ vs 
dimension $d$.  The circles represent the exact values calculated by diagonalizing  
$H$ analytically or numerically. The values of the asymptotic formula 
Eq.~(\ref{eq_h_asym})  are plotted by triangles.}
\label{fig:h_vs_dim}
\end{figure}

\begin{figure}
\includegraphics[width=\linewidth]{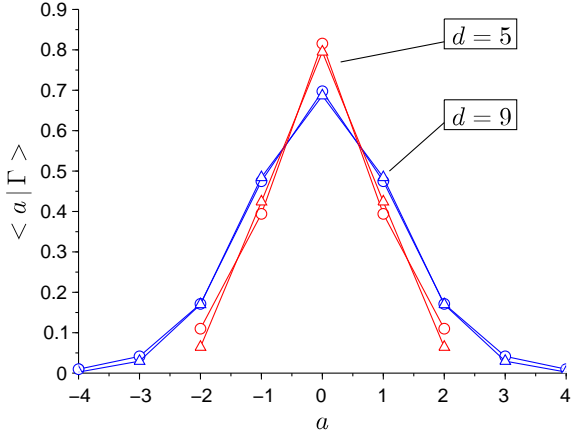}
\caption{(Color online) The components of  the minimum-uncertainty state 
$\ket{\Gamma}$. The components in the position basis 
$\{\ket{a},\   -(d-1)/2 \le a \le (d-1)/2\}$ are plotted vs $a$ for $d=5$ and $d=9$ 
cases.  The circles are the values obtained by numerical calculations. The triangles 
represent the values by the asymptotic form of Eq.~(\ref{eq_alpha_asym}). 
The normalization constants $\cal N$ are determined numerically. 
}
\label{fig:gama}
\end{figure}

In Fig.~\ref{fig:h_vs_dim} we compare the exact values of $h$ with those 
obtained by the asymptotic formula Eq.~(\ref{eq_h_asym}).  This shows that 
the asymptotic form is already a rather good approximation for relatively low 
dimensions.  The components of the minimum-uncertainty state $\braket{a|\Gamma}$, 
the values by numerical calculation and by the asymptotic 
form Eq.~(\ref{eq_alpha_asym}),  are plotted in Fig.~\ref{fig:gama}. 
We see  that the asymptotic form provides an unexpectedly good approximation 
even for the $d=5$ case.

We briefly sketch how the inequality (\ref{eq_main_inequality}) of the 
certainty $C(\phi)$ is reduced to the usual uncertainty relation of the position and momentum variables  in the continuum limit. First we analyze the expectation value 
$\braket{\phi|Q|\phi}$. In the continuum limit, the summation over $a$ becomes 
an integral over $x$, and the exponential function 
$\exp(i\frac{2\pi}{d}a)=\exp(i\frac{2\pi}{L}x)$ can be expanded. 
Thus we have
\begin{align}
   & \braket{\phi|Q|\phi} = \sum_a e^{i\frac{2\pi}{d}a} |c_a|^2 
                                                            \nonumber \\
    &= 1 +i\frac{2\pi}{L}\braket{\hat x}
              -\frac{1}{2}\left(\frac{2\pi}{L}\right)^2 \braket{\hat x^2}
                                                             +  O\left( \frac{1}{d^{3/2}} \right), 
\end{align}
where
\begin{align}
   & \braket{\hat x} = \int\!\! dx \phi^*(x) x \phi(x),   \nonumber \\
   & \braket{\hat x^2} = \int\!\! dx \phi^*(x) x^2 \phi(x). \nonumber
\end{align}
The modulus of $\braket{\phi|Q|\phi}$ then takes the form 
\begin{align}
  |\braket{\phi|Q|\phi}| = 1 - \frac{\pi}{\sigma^2 d} (\Delta x)^2
                   +  O\left( \frac{1}{d^{3/2}} \right),
\end{align} 
in terms of the standard deviation of position coordinate 
defined by $\Delta x = \sqrt{  \braket{\hat x^2}-\braket{\hat x}^2}$.
Similarly, $|\braket{\phi|P|\phi}|$ is expressed as
\begin{align}
   |\braket{\phi|P|\phi}| =  1 - \frac{\pi \sigma^2}{d} (\Delta p)^2
                     + O\left( \frac{1}{d^{3/2}} \right),
\end{align}
where $\Delta p$ is the usual standard deviation of momentum coordinate. 
Meanwhile, the asymptotic form of $h$ has already been 
obtained in Eq.~(\ref{eq_h_asym}). 
Combining all these results, we find that the inequality (\ref{eq_main_inequality})
of the certainty $C(\phi)$ is reduced to 
\begin{align}
     \frac{1}{\sigma^2} (\Delta x)^2 + \sigma^2 (\Delta p)^2 \ge 1,
                          \label{eq_continuum_inequality}
\end{align}
in the leading order of $1/d$. 

It is evident that, for a given wave function $\phi(x)$, 
the scale factor $\sigma$ is arbitrary, since $\sigma$ is a sort of artifact in the 
procedure of the continuum-limit scheme. The left-hand side of 
the above inequality (\ref{eq_continuum_inequality}) takes the minimum 
value $2 \Delta x \Delta p$ when $\sigma = \sqrt{\Delta x/\Delta p}$. 
Thus we arrive at the usual uncertainty relation of the position and momentum 
variables.  
\begin{align}
       \Delta x \Delta p \ge \frac{1}{2}.
\end{align}

\smallskip
\section{Non-negative quasi probability and its optimality} 
\label{sec_non-negative}
The minimum-uncertainty states are defined as 
\begin{align}
    \ket{\alpha,\beta} = P^\alpha Q^\beta \ket{\Gamma},
                \ (\alpha,\beta=0,1,\ldots,d-1).
\end{align}
The position and momentum distributions of $\ket{\alpha,\beta}$ are given by 
\begin{align}
   |\braket{a|\alpha,\beta}|^2 &= \Gamma_{a-\alpha}^2, \\
   |\braket{\tilde b|\alpha,\beta}|^2 &= \Gamma_{b-\beta}^2.
\end{align}
Note that $\Gamma_a \equiv \braket{a|\Gamma}$ has a peak at $a=0$, 
which can be seen from the analytical results in the low-dimensional cases 
and the numerical results  for higher dimensions. 
Therefore, the position and momentum distribution of $\ket{\alpha,\beta}$ 
have a peak at $a=\alpha$ and $b=\beta$, respectively. 

The $d^2$ minimum-uncertainty states $\ket{\alpha,\beta}$ are not mutually  orthogonal, but they comprise an overcomplete set in the state vector space 
$\mathbb{C}^d$. 
The completeness relation of $\ket{\alpha,\beta}$ takes the form
\begin{align}
   \frac{1}{d}\sum_{\alpha,\beta=0}^{d-1} 
                    \ket{\alpha,\beta}\bra{\alpha,\beta} =  \mbold{1}.
                           \label{eq_minimum_completeness}
\end{align}
In order to derive this completeness relation, we employ the 
following useful identity which holds for any operator $\Omega$:
\begin{align}
  \sum_{\alpha,\beta=0}^{d-1}\omega^{\alpha b-\beta a} 
                        P^{\alpha}Q^{\beta} \Omega Q^{-\beta}P^{-\alpha} 
   = d \tr[Q^{-b}P^{-a}\Omega] P^{a}Q^b,
                               \label{eq_omega_identity1}
\end{align}
or equivalently, 
\begin{align}
   & P^{\alpha}Q^{\beta} \Omega Q^{-\beta}P^{-\alpha}
                        \nonumber \\  
 & = \frac{1}{d}\sum_{a,b=0}^{d-1}\omega^{-\alpha b+\beta a} 
          \tr[Q^{-b}P^{-a}\Omega] P^{a}Q^b.
                            \label{eq_omega_identity2}
\end{align}
This identity can be obtained by using  the commutation relation $QP=\omega PQ$ 
together with the mutual orthogonality and completeness of the set of operators 
$\{P^{\alpha}Q^\beta\}_{\alpha,\beta=0}^{d-1}$ in the operator space. 
Setting $a=b=0$ and $\Omega=\ket{\Gamma}\bra{\Gamma}$ in the above 
identity (\ref{eq_omega_identity1}), we obtain the completeness of 
the minimum-uncertain states (\ref{eq_minimum_completeness}).

Based on these observations, it is reasonable to define the 
quasi probability distribution $D(\alpha,\beta)$  for a given state $\rho$ with respect to the position and momentum coordinates $\alpha$ and $\beta$ as 
follows: 
\begin{align}
   D(\alpha,\beta) \equiv 
           \frac{1}{d}\braket{\alpha,\beta|\rho|\alpha,\beta} 
         = \tr[\rho \Delta(\alpha,\beta)],
\end{align}
where we introduced the phase point operator $\Delta(\alpha,\beta)$ 
given by
\begin{align}
   \Delta(\alpha,\beta) = \frac{1}{d} \ket{\alpha,\beta}\bra{\alpha,\beta}.
\end{align}
Note that $D(\alpha,\beta)$ is non-negative and normalized to unity when 
summed over all phase space points $(\alpha,\beta)$. However, the states 
$\ket{\alpha,\beta}$ are not mutually orthogonal, and therefore distinct 
phase space points $(\alpha,\beta)$ are not regarded as exclusive events.  
This is the reason why we call $D(\alpha,\beta)$ a {\it quasi} probability distribution. 

The phase point operator $\Delta(\alpha,\beta)$ satisfies the following 
relations if summed over $\alpha$ or $\beta$:
\begin{align}
  \sum_{\beta=0}^{d-1} \Delta(\alpha,\beta) 
        &= \sum_{a=0}^{d-1} \Gamma_{a-\alpha}^2\ket{a}\bra{a},     
                                  \label{eq_margin_a} \\
  \sum_{\alpha=0}^{d-1} \Delta(\alpha,\beta) 
        &= \sum_{b=0}^{d-1} \Gamma_{b-\beta}^2\ket{\tilde b}\bra{\tilde b}.
                                  \label{eq_margin_b}
\end{align}
The first equation (\ref{eq_margin_a}) can be obtained by summing over 
$\beta$ in  Eq.~(\ref{eq_omega_identity2}) with 
$\Omega=\ket{\Gamma}\bra{\Gamma}$. Similarly the second equation   (\ref{eq_margin_b}) also follows from Eq.~(\ref{eq_omega_identity2}). 
These relations (\ref{eq_margin_a}) and (\ref{eq_margin_b}) imply that  
the quasi probability distribution $D(\alpha,\beta)$ has the following 
marginal distributions: 
\begin{align}
   \sum_{\beta=0}^{d-1} D(\alpha,\beta) 
        &= \sum_{a=0}^{d-1} \Gamma_{a-\alpha}^2\braket{a|\rho|a},     
                                  \label{eq_Dmargin_a} \\
  \sum_{\alpha=0}^{d-1} D(\alpha,\beta) 
        &= \sum_{b=0}^{d-1} \Gamma_{b-\beta}^2\braket{\tilde b|\rho|\tilde b}.
                                  \label{eq_Dmargin_b}
\end{align}
We find that the marginal distributions are smeared out in the sense  
that $D(\alpha,\beta)$ summed over $\beta$, for example,  
gives the weighted average of $\braket{a|\rho|a}$ with the weight  
centered at $a=\alpha$. 

It is evident that the phase  point operator $\Delta(\alpha,\beta)$ 
respects the translational covariance, 
\begin{align}
    P^aQ^b \Delta(\alpha,\beta) Q^{-b}P^{-a} = \Delta(\alpha+a,\beta+b),
\end{align}
which implies that if $D(\alpha,\beta)$ is the quasi probabilities of a state $\rho$ 
then the quasi probabilities of  $\rho' = P^aQ^b \rho  Q^{-b}P^{-a}$ is given by
$D(\alpha-a,\beta-b)$. The phase point operator is also 
covariant under the Fourier transformation; that is, 
$F \Delta(\alpha,\beta) F^\dagger = \Delta(-\beta,\alpha)$, but 
not covariant under the more general symplectic transformation considered in 
\cite{Horibe2002,Gross2006_7,Horibe2013}.

For the odd-dimensional system, the Wigner function of  
Wootters \cite{Wootters1987} and Cohendet {\it et al}. \cite{Cohendet1988}  
is defined as 
$D_W(\alpha,\beta) = \tr[\rho \Delta_W(\alpha,\beta)]$ with
the phase point operator given by
\begin{align}
    \Delta_{{\rm W}}(\alpha,\beta) = 
                 \frac{1}{d}P^\alpha Q^\beta T Q^{-\beta}P^{-\alpha}.              
\end{align}
This Wigner function has sharp marginal distributions since 
\begin{align}
  \sum_{\beta=0}^{d-1} \Delta_W(\alpha,\beta) 
        &= \ket{\alpha}\bra{\alpha},     
                       \label{eq_Wmargin_a}  \\
  \sum_{\alpha=0}^{d-1} \Delta_W(\alpha,\beta) 
        &= \ket{\tilde \beta}\bra{\tilde \beta}.
                       \label{eq_Wmargin_b}
\end{align}
However, the Wigner functions $D_W(\alpha,\beta)$ may take negative values, and 
they are non-negative only for special states called stabilizer states  \cite{Gross2006_7}, 
since $\Delta_W(\alpha,\beta)$ is not positive semidefinite. 
Using the mutual orthogonality and completeness of $\Delta_W(\alpha,\beta)$ in 
the operator space, we can easily express $\Delta(\alpha,\beta)$ in terms of 
$\Delta_W(\alpha,\beta)$. The result is given by
\begin{align}
   \Delta(\alpha,\beta) =  \sum_{\alpha',\beta'=0}^{d-1} 
                  w(\alpha-\alpha',\beta-\beta') \Delta_{{\rm W}}(\alpha',\beta'),
\end{align}
where
\begin{align}
  w(\alpha,\beta) = \braket{\Gamma|
                      \Delta_{{\rm W}}(\alpha,\beta)| \Gamma}.
\end{align}
We see that the phase point operator $\Delta(\alpha,\beta)$ built with the 
minimum-uncertainty states can be written in the form of convolution of the 
weight $w(\alpha,\beta)$ and $\Delta_W(\alpha,\beta)$, and thus it acquires 
nonnegativity at the cost of losing the sharp marginal property. 

The quasi probability distribution based on the minimum-uncertainty states is 
non-negative, but its marginal distributions are smeared out, as shown in 
Eqs.~(\ref{eq_Dmargin_a}) and (\ref{eq_Dmargin_b}). 
A natural question is whether there exists a non-negative quasi probability 
distribution which satisfies sharper marginal conditions. 
In what follows, we show that the answer is ``no'' as long as  the translational covariance in the phase space is assumed.

Let $\Lambda(\alpha,\beta)$ be phase point operators of a non-negative 
quasi probability distribution with the translational covariance.  
To quantify the sharpness of the marginal distributions, we define 
\begin{align}
  \sigma & \equiv \left| 
      \tr\left[\sum_{\beta=0}^{d-1} \Lambda(\alpha,\beta) Q \right] 
                         \right|,   \\ 
    \tau   & \equiv \left| 
      \tr\left[\sum_{\alpha=0}^{d-1} \Lambda(\alpha,\beta) P \right] 
                         \right|.
\end{align}
Because of the translational covariance, $\sigma$ and $\tau$ are  
independent of $\alpha$ and $\beta$, respectively. In the case of 
$\Delta_W(\alpha,\beta)$ by Wootters and Cohendet {\it et al}., 
we find that $\sigma=\tau=1$ since the marginal 
conditions are perfectly sharp as shown in 
Eqs.~(\ref{eq_Wmargin_a},\ref{eq_Wmargin_b}). 
However, for $\Delta(\alpha,\beta)$ based on the minimum-uncertainty states, 
we have $\sigma=\tau=h$, which is less than 1 if $d \ge 2$.

The translational covariance implies that $\Lambda(\alpha,\beta)$ can be 
written as 
\begin{align}
   \Lambda(\alpha,\beta) 
      = \frac{1}{d} P^\alpha Q^\beta K Q^{-\beta}P^{-\alpha},
\end{align}
where $K=d\Lambda(0,0)$ is a Hermitian operator with $\tr K=1$ since $\Lambda(\alpha,\beta)$ should be Hermitian and normalized as 
$\sum_{\alpha,\beta=0}^{d-1} \Lambda(\alpha,\beta) = 1$. 
In addition, $K$ should be positive semidefinite to ensure that the quasi probabilities 
are non-negative.  Thus $K$ can be regarded as a state on $\mathbb{C}^d$.
In terms of $K$, the measures of sharpness, $\sigma$ and $\tau$, take the following 
simple form:
\begin{align}
   \sigma = |\tr[KQ]|,\ \tau = |\tr[KP]|. 
\end{align} 

Here it should be noticed that the theorem in Sec.~\ref{sec_Minimum}
holds also for mixed states; that is, for any state $\rho$, we have 
\begin{align}
    \left| \tr[\rho Q] \tr[\rho P] \right| \le h^2, 
\end{align}
where the equality holds if and only if $\rho=\ket{\alpha,\beta}\bra{\alpha,\beta}$.
This can be shown by the following inequalities:
\begin{align}
  &\left| \tr[\rho Q] \tr[\rho P] \right|^{1/2} 
    \le \frac{1}{2} \left( | \tr[\rho Q] | +  |\tr[\rho P]| \right)
                              \nonumber \\
  &\le \sum_i r_i \frac{1}{2} \left(
            |\braket{\phi_i|Q|\phi_i}| + |\braket{\phi_i|P|\phi_i}| 
                                                           \right) \le h,
\end{align}
where we used the spectral decomposition 
$\rho =  \sum_i r_i \ket{\phi_i}\bra{\phi_i}$. 

Using this extended theorem, we obtain 
\begin{align}
     \sigma \tau \le h^2,
\end{align} 
where the equality holds if and only if 
$K= \ket{\alpha_0,\beta_0}\bra{\alpha_0,\beta_0}$ with 
$\alpha_0,\beta_0=0,1,\ldots,d-1$.
This implies that the upper bound of the sharpness $\sigma\tau$ 
is attained by $\Lambda(\alpha,\beta)=\Delta(\alpha+\alpha_0,\beta+\beta_0)$.  
Thus we conclude that the quasi probability distribution based on 
$\Delta(\alpha,\beta)$ is optimal and unique up to a cyclic relabeling of the position 
and momentum coordinates; $\alpha \rightarrow \alpha+\alpha_0$ and 
$\beta \rightarrow \beta+\beta_0$. 
                                             
\section{Completeness}  \label{sec_completeness}
It is desirable that the quasi probability distribution completely determines 
the state of the system. This requires  that the set of phase point operators 
$\{\Delta(\alpha,\beta)\}_{\alpha,\beta=0}^{d-1}$ should  be 
complete in the operator space.  To see this, we calculate the Fourier 
transform of $\Delta(\alpha,\beta)$.
\begin{align}
  \tilde\Delta(m,n) & \equiv \frac{1}{d} \sum_{\alpha,\beta=0}^{d-1}
                                        \omega^{\alpha n-\beta m} \Delta(\alpha,\beta) 
                                               \nonumber \\
        & = \frac{1}{d} \braket{\Gamma | Q^{-n}P^{-m}| \Gamma}P^mQ^n
                                              \nonumber \\
       & = \frac{1}{d} \braket{\Gamma | Q^{n}P^{m}| \Gamma}P^mQ^n.
\end{align}
We employed Eq.~(\ref{eq_omega_identity1}) with 
$\Omega = \ket{\Gamma}\bra{\Gamma}$ to obtain the second line of 
the above equation, and the reflection symmetry 
$T\ket{\Gamma} = \ket{\Gamma}$ was also used in the last line. 
Since the set of operators  
$\{P^mQ^n\}_{m,n=0}^{d-1}$ is complete, the completeness of 
the phase point operators is equivalent to the conditions given by 
\begin{align}
    f_{m,n} \equiv 
       \braket{\Gamma | P^{m}Q^{n}| \Gamma} \ne 0,\ (m,n=0,1,\ldots,d-1).
                    \label{eq_complete_condition}
\end{align}
$f_{mn}$ has the following symmetries: 
\begin{align}
   f_{mn} &= f_{-m,-n},   \nonumber \\
   f_{mn} &= f_{nm}, \nonumber \\
   f_{mn} &= \omega^{-mn}f_{m,-n}, \nonumber \\
   f_{mn} & = \omega^{-mn}f_{mn}^*. \nonumber
\end{align}
We used the fact that  the state $\ket{\Gamma}$ is invariant under 
the Fourier transformation, and components $\braket{a|\Gamma}$ can 
be taken to be real values. 

Here we have different results depending on whether the dimension $d$ 
is even or odd. 
When $d$ is even, some of the conditions (\ref{eq_complete_condition}) are 
clearly violated.  For example, we find that 
\begin{align}
     \braket{\Gamma | P^{d/2}Q^{d/2} |\Gamma}
       = \frac{1}{2h} \braket{\Gamma |\{P^{d/2}Q^{d/2},H\} |\Gamma}=0,
\end{align}
since the operator $P^{d/2}Q^{d/2}$ anticommutes with $H$. 
Using the symmetries of $f_{mn}$, we also observe that 
\begin{subequations}
\begin{align}
    & \braket{\Gamma | P^mQ^{d/2} |\Gamma} = 0,\ (m=\text{odd}),
                           \\
    & \braket{\Gamma | P^{d/2}Q^{n} |\Gamma} = 0,\ (n=\text{odd}).
\end{align}
\end{subequations}
We remark that it is only in those cases that 
$\braket{\Gamma | P^mQ^n |\Gamma}$ vanishes, which can be shown by 
an analysis similar to the one in the odd-dimensional case given later 
in this section. 
Thus, the phase point operators $\Delta(\alpha,\beta)$ are not 
complete if $d$ is even. Let us examine the qubit ($d=2$) case more closely. 
In this case we can write the phase point operator as 
\begin{align}
    \Delta(\alpha,\beta) = \frac{1}{4} 
               \left( 1+\mbold{n}^{(\alpha,\beta)}\cdot \mbold{\sigma} 
                   \right), \  (\alpha,\beta=0,1), 
\end{align} 
where the Bloch vectors $\mbold{n}^{(\alpha,\beta)}$ are given in 
Eq.~(\ref{eq_Bloch}).  Since the $y$-components of 
$\mbold{n}^{(\alpha,\beta)}$ are 0, the set of $\Delta(\alpha,\beta)$'s  is not 
complete in the whole qubit space. However, it is interesting that it is still complete in the qubit space of real amplitudes. 

When $d$ is odd, on the other hand, the conditions 
(\ref{eq_complete_condition}) are  satisfied: the set of phase point 
operators $\Delta(\alpha,\beta)$ is complete, which will be shown in the 
rest of this section. 

\subsection{Equations for $\braket{\Gamma|P^mQ^n|\Gamma}$}
In this subsection, we will derive some equations fulfilled by 
$f_{mn} \equiv \braket{\Gamma|P^mQ^n|\Gamma}$.
Here, the dimension $d$ is arbitrary (odd or even).

We begin with the following two evident equations:  
\begin{subequations} 
\begin{align}
   \braket{\Gamma|HP^mQ^n|\Gamma} &= h  \braket{\Gamma|P^mQ^n|\Gamma}, 
                  \\
   \braket{\Gamma|P^mQ^nH|\Gamma} &= h  \braket{\Gamma|P^mQ^n|\Gamma},
\end{align}
\end{subequations}
and we write them in terms of $f_{mn}$ as 
\begin{subequations} \label{eq_eq_fmn}
\begin{align}
   & \frac{1}{4} \left( 
        f_{m+1,n} + f_{m-1,n} + \omega^mf_{m,n+1} + \omega^{-m}f_{m,n-1} 
                       \right)= hf_{m,n},   \\
   & \frac{1}{4} \left( 
        \omega^nf_{m+1,n} + \omega^{-n}f_{m-1,n} + f_{m,n+1} + f_{m,n-1} 
                       \right)= hf_{m,n}.
\end{align}
\end{subequations}

Regarding  $f_{m,n}$ as the  $(m,n)$-entry of the vector 
$\ket{f}$ in $\mathbb{C}^d \otimes \mathbb{C}^d$, we write 
Eqs.~(\ref{eq_eq_fmn}) in the form 
\begin{subequations} 
\begin{align}
  {\cal H}_L \ket{f} &= h\ket{f},\\
  {\cal H}_R \ket{f} &= h\ket{f},
\end{align}
\end{subequations}
where 
\begin{subequations}
\begin{align}
   {\cal H}_L &= \frac{1}{4} 
        \left(  P^{-1} \otimes \mbold{1} + P \otimes \mbold{1}
              + Q \otimes P^{-1} + Q^{-1} \otimes P \right), \\
   {\cal H}_R &= \frac{1}{4} 
        \left( P^{-1} \otimes Q + P \otimes Q^{-1}
              +\mbold{1} \otimes P^{-1} + \mbold{1} \otimes P \right). 
\end{align}
\end{subequations}
Thus, $\ket{f}$ is a simultaneous eigenstate of  
${\cal H}_L$ and ${\cal H}_R$ with eigenvalue $h$. 

Let us see ${\cal H}_L$ more closely. 
Express the space $\mathbb{C}^d \otimes \mathbb{C}^d$ as 
$\bigoplus_{b=0}^{d-1} V^{(b)}$, where 
\begin{align}
  & V^{(b)} \equiv {\rm Span}\{ \ket{\Psi^{(b)}_{a}},\ a=0,\ldots,d-1 \}, \\
  & \ket{\Psi^{(b)}_{a}} \equiv \ket{a-b} \otimes \ket{\tilde b}. 
\end{align}
We then observe that each term in ${\cal H}_L$ transforms the states 
$ \ket{\Psi^{(b)}_{a}}$ in the following way: 
\begin{align}
   P^{-1} \otimes \mbold{1} \ket{\Psi^{(b)}_a} &= \ket{\Psi^{(b)}_{a-1}}, 
           \nonumber \\
   P \otimes \mbold{1} \ket{\Psi^{(b)}_a} &= \ket{\Psi^{(b)}_{a+1}}, 
           \nonumber \\ 
   Q \otimes P^{-1} \ket{\Psi^{(b)}_a} &= \omega^a \ket{\Psi^{(b)}_{a}},  
           \nonumber \\
   Q^{-1} \otimes P \ket{\Psi^{(b)}_a} &= \omega^{-a} \ket{\Psi^{(b)}_{a}}. 
           \nonumber
\end{align}
This implies that ${\cal H}_L = \bigoplus_{b=0}^{d-1} H$, and 
$h$ is the maximum eigenvalue of ${\cal H}_L$, which is $d$-fold degenerate. 
The same thing is true for ${\cal H}_R$．Therefore, the maximum eigenvalue of 
${\cal H}_L+{\cal H}_R$ is $2h$, and $\ket{f}$ is one of the associated 
eigenstates. Thus we obtain 
\begin{align}
   ( {\cal H}_L+{\cal H}_R) \ket{f} = 2h\ket{f}. 
        \label{eq_eigen_f}
\end{align}

It is useful to define real quantities $g_{mn}$ as 
\begin{align}
   g_{mn} \equiv e^{\frac{2\pi i}{d}\frac{mn}{2}} f_{mn}. 
\end{align}
$g_{mn}$ is real and has the following symmetries:
\begin{align}
   & g_{mn}^* = g_{mn},\ g_{mn}=g_{nm}, \nonumber \\
   & g_{m,n}=g_{-m,n}=g_{n,-m}.  \nonumber
\end{align}
Note that $g_{mn}$ is not periodic with period $d$ for $m$ and $n$, 
rather satisfies the following relations:
\begin{align}  
   g_{m\pm d,n} = (-)^n g_{m,n},\ g_{m,n \pm d} = (-)^m g_{m,n}. 
             \nonumber 
\end{align}


\subsection{$\braket{\Gamma|P^mQ^n|\Gamma} \ne 0 $ when $d$ is odd}
\label{sec_odd_d}
In this subsection, we assume that $d$ is odd, and we fix the range of 
the indices $m,n,m',n'$ as 
\begin{align}
    -(d-1)/2 \le m,n,m',n' \le (d-1)/2. \label{eq_oddrange}
\end{align}
We will show that $g_{mn}$ are strictly positive. 
Rewriting Eq.~(\ref{eq_eigen_f}), we obtain the eigen equation for 
$\ket{g} \in \mathbb{C}^d \otimes \mathbb{C}^d$ with 
$\braket{mn|g}=g_{mn}$.
\begin{align}
    {\cal K} \ket{g} = 2h \ket{g}, 
\end{align}
where
\begin{align}
     {\cal K}_{mn,m'n'} = e^{\frac{2\pi i}{d}\frac{mn}{2}} 
                               ({\cal H}_L + {\cal H}_R)_{mn,m'n'} 
                              e^{-\frac{2\pi i}{d}\frac{m'n'}{2}}.
                              \nonumber
\end{align}
We find that ${\cal K}_{mn,m'n'}$ is given by 
\begin{align}
   {\cal K}_{mn,m'n'} = &
       \frac{1}{2}D_{mm'}[(-1)^n] \cos\left( \frac{\pi n}{d} \right) 
                \delta_{nn'}                        \nonumber \\ 
     + &\frac{1}{2}\cos\left( \frac{\pi m}{d} \right)
                \delta_{mm'} D_{nn'}[(-1)^m],       \nonumber 
\end{align}
where the $d\times d$ matrix $D(\sigma)$ is defined as 
\begin{align}
   D_{mm'}(\sigma) = & \delta_{|m-m'|,1}
        +\sigma \big(\delta_{m,(d-1)/2}\delta_{m',-(d-1)/2} 
                                          \nonumber \\
                     & +  \delta_{m,-(d-1)/2}\delta_{m',(d-1)/2} \big), 
                          \nonumber
\end{align} 
or 
\begin{align}
 D(\sigma) =\left[ 
      \begin{array}{ccccccc}
              0          &   1         & 0         & \cdots &  0         & \sigma \\
              1          &   0         & 1         & \ddots &  0         & 0          \\ 
              0           &  1         & \ddots & \ddots & \ddots & \vdots   \\ 
              \vdots   & \ddots & \ddots & \ddots & 1         & 0            \\ 
              0           &  0         & \ddots & 1          & 0         & 1           \\ 
              \sigma  & 0         & \cdots  & 0          & 1         & 0           \\ 
      \end{array} 
                  \right].   \nonumber
\end{align} 
Note that $\cos\left( \frac{\pi n}{d} \right)$ and $\cos\left( \frac{\pi m}{d} \right)$
are positive since the ranges of $m,n$ are given by Eq.~(\ref{eq_oddrange}). 
However, the Perron-Frobenius theorem is not yet applicable to 
${\cal K}$, since the nondiagonal elements may be negative depending on 
the even-oddness of $m,n$. 

Here, we notice that ${\cal K}$ has the following reflexion symmetries:
\begin{align}
   {\cal K}_{m,n,m',n'} = {\cal K}_{-m,n,-m',n'}={\cal K}_{m,-n,m',-n'}. 
\end{align}
And $g_{mn}$ is also symmetric under these reflexions. 
To exploit this fact, we rewrite the eigen equation 
${\cal K} \ket{g} = 2h\ket{g}$ 
in the base $\{\ket{e_{a,b}},\ a,b=0,1,\ldots,(d-1)/2\}$ 
which respects the reflexion symmetries. 
\begin{align}
     \ket{e_{ab}} & \equiv \ket{e_a} \otimes \ket{e_b}, \\
     \ket{e_a} & \equiv \frac{1}{\sqrt{2(1+\delta_{a0})}} 
        \left( \ket{a} + \ket{-a}  \right).
\end{align} 
In this base, the eigen equation reads 
\begin{align}
    {\cal K}^{(S)} g^{(S)} = 2h g^{(S)},
\end{align}
where
\begin{align}
    g^{(S)}_{ab} = \braket{e_{ab}|g}, 
\end{align}
and 
\begin{align}
   {\cal K}^{(S)}_{ab,a'b'} =& \braket{ e_{ab} |{\cal  K}| e_{a'b'}} 
                               \nonumber \\ 
   =& \frac{1}{2}D^{(S)}_{aa'}[(-1)^b] \cos
           \left( \frac{\pi b}{d} \right) \delta_{bb'}
                               \nonumber \\
      &+\frac{1}{2}\cos
           \left( \frac{\pi a}{d} \right)\delta_{aa'} D^{(S)}_{bb'}[(-1)^a] .
\end{align}
Here, the $(d+1)/2 \times (d+1)/2$ matrix $D^{(S)}(\sigma)$ is given by 
\begin{align}
   D^{(S)}_{aa'}(\sigma) &= \braket{ e_a | D(\sigma) | e_{a'}} 
                              \nonumber \\
       &= \sqrt{1+\delta_{a,0}}\delta_{a,a'-1} 
         + \sqrt{1+\delta_{a',0}}\delta_{a',a-1} 
                             \nonumber \\
        & + \sigma \delta_{a,(d-1)/2}\delta_{a',(d-1)/2}, 
                             \nonumber
\end{align}
or 
\begin{align}
 D^{(S)}(\sigma) =\left[ 
      \begin{array}{ccccccc}
          0              & \sqrt{2} & 0         & \cdots   & 0         \\
          \sqrt{2}  &  0            & 1          & \ddots  & \vdots    \\ 
          0              &  1            & \ddots & \ddots   & 0        \\ 
          \vdots     & \ddots    & \ddots  & 0            & 1           \\ 
          0             & \ldots      & 0        & 1           & \sigma \\ 
      \end{array}  
                  \right]. \nonumber 
\end{align} 

Now we examine the real symmetric matrix ${\cal K}^{(S)}+\mbold{1}$. 
All elements are non-negative, and the diagonal elements are strictly positive. 
Further, the matrix elements $({\cal K}^{(S)}+\mbold{1})_{ab,a'b'}$ 
are strictly positive if the points $(a,b)$ and $(a',b')$ are the nearest neighbors on the two-dimensional integer lattice.
Therefore, all elements of 
$A \equiv ({\cal K}^{(S)}+\mbold{1})^{d-1}$ are strictly positive. 
Evidently, $g^{(S)}$ is the eigen vector of $A$ with the maximum eigenvalue 
$(2h+1)^{d-1}$.  
According to the theorem of Perron-Frobenius, all components of $g^{(S)}$ 
can be taken to be strictly positive. 
This further implies that all $g_{mn}$ are strictly positive 
because of the reflection invariance of $g_{mn}$. 
Thus we have shown that $\braket{\Gamma|P^mQ^n|\Gamma} \ne 0 
$ when $d$ is odd. 

\section{Summary and concluding remarks}   \label{sec_conclusion}
The aim of this paper is to construct the minimum-uncertainty states and 
the non-negative quasi probability distribution for a qudit. They are 
the finite-dimensional counter parts of the coherent states and the Husimi 
function of the continuous quantum mechanics.  

We reexamined the theorem of Massar and Spindel for the uncertainty relation of the two unitary operators related by the discrete Fourier transformation, and we showed that some assumptions in their proof can 
be justified if we use the Perron-Frobenius theorem.  
The minimum-uncertainty states are the ones that  
saturate this uncertainty inequality. 
By introducing a scale factor in the continuum limit, we showed that they  
approach the coherent states with different widths.  

We constructed the non-negative quasi probability distribution, 
of which marginal distributions are smeared out as in the Husimi function. 
However, this quasi probability distribution is shown to be optimal in 
the sense that there does not exist a non-negative and translationally 
covariant quasi probability distribution with sharper marginal properties. 
Generally, the completeness is one of the desirable properties of a quasi probability distribution; that is, it contains full information of the state.  
We showed that the obtained quasi probability is indeed complete if the 
dimension of the state space is odd, whereas it is unfortunately not 
if the dimension is even.
It is well known that the Wigner function in the even-dimensional 
case is much more involved than in the odd-dimensional case 
(see, e.g., \cite{Leonhardt1995_6,Takami2001}). 
Further investigation for this even-odd issue of quasi probabilities is 
certainly needed. 

The Wigner function may take negative values. 
In Refs. \cite{Cohendet1988,Hashimoto2007}, however, it is shown that 
one can define non-negative quasi probabilities by introducing an auxiliary variable into the Wigner function, and solve the dynamics of a quantum system  stochastically. It will be of interest in future studies to apply our quasi probability distribution to this line of research.


\end{document}